\pdfoutput=1
\documentclass{JINST}

 \usepackage[pdftex]{graphics}
\usepackage[pdftex]{graphicx}

\usepackage{amssymb}
\usepackage{lineno}
\usepackage{lipsum}

\linenumbers 

\title{PHIL photoinjector test line\thanks{Web site of the project :http://phil.lal.in2p3.fr/}}

\author{M. Alves$^a$, C. Arnault$^a$, D. Auguste$^a$, J.L. Babigeon $^a$ 
 F. Blot$^a$, J. Brossard$^a$, C. Bruni\thanks{Corresponding
author: bruni@lal.in2p3.fr}, S. Cavalier$^a$, J.N. Cayla$^a$, V. Chaumat$^a$, J. Collin$^a$, M. Dehamme$^a$, M. Demarest$^a$, J.P. Dugal$^a$, M. Elkhaldi$^a$, I. Falleau$^a$, A. Gonnin$^a$, M. Jore$^a$, E. Jules$^a$, B. Leluan$^a$, P. Lepercq$^a$, F. Letellier$^a$, E. Mandag$^a$, J.C. Marrucho$^a$, B. Mercier$^a$, E. Mistretta$^a$, C. Prevost$^a$, R. Roux$^a$, V. Soskov$^a$, A. Toutain$^a$, A. Variola$^a$, O Vitez$^a$, H. Monard$^a$\thanks{Project manager: monard@lal.in2p3.fr}\\
\llap{$^a$}Laboratoire de l'acc\'el\'erateur lin\'eaire,\\
  Universit\'e Paris-Sud 11, UMR 8607, B\^atiment 200, 91898 Orsay
  Cedex, France\\
  E-mail: \email{bruni@lal.in2p3.fr}}

\abstract{LAL is now equiped with its own platform for photoinjectors tests and Research and Developement, named PHIL (PHotoInjectors at LAL). This facility has two main purposes: push the limits of the photoinjectors performances working on both the design and the associated technology and  provide a low energy (MeV) short pulses (ps) electron beam for the interested users. Another very important goal of this machine will be to provide an opportunity to form  accelerator physics students, working in a high technology environment
To achieve this goal a test line was realised equipped with an RF source, magnets and beam diagnostics. In this article we will desrcibe the PHIL beamline and its characteristics together with the description of the first two photoinjector realised in LAL and tested: the ALPHAX and the PHIN RF Guns. }


\begin{document}

\section{Introduction}
For many years LAL (Laboratoire de L'Acc\'el\'erateur Lin\'eaire at Orsay, France)
has built RF guns for different projects starting with CANDELA \cite{CAND} at Orsay, France going to
ALPHAX for Stratchlyde university \cite{STRAT} in the UK, ELYSE \cite{ELYSE} for the Laboratoire de Chimie-Physique at Orsay and recently for the probe beam and the test beam at CERN/CTF3 \cite{ref2, ref3, EPAC06brossard}. At present LAL is equipped with its own photoinjector test line named PHIL (PHotoInjector at LAL) proposed in the European Community REsearch Infrastucture Activity CARE \cite{PHIL} \footnote{We acknowledge 
the support of the European Community-Research Infrastructure Activity under the FP6 "Structuring the European Research Area" 
programme (CARE, contract number RII3-CT-2003-506395), http:/www.infn.it/phin} able to work with several RF guns at 3 GHz.

PHIL was designed for two main purposes :

\begin{itemize}
	\item research and development on the electron sources,
	\item providing the beam to users interested in a low energy (9 MeV), low emittance short pulse (5 ps) electron beam.
\end{itemize}

The research and development program for PHIL is driven by the will to obtain high gradient electric field up to 100 MV/m, subpicosecond electron beam duration, photocathodes with longer lifetime, a lower dark current, a better understanding of the electron beam dynamics inside the gun, the comparison between different RF guns and cathodes, and pushing the limits of the number of cells for an RF gun \cite{PAC2011}, etc...  PHIL will also be a platform for welcoming users interested in a low emittance, well defined energy, low energy electron beam. For example measuring the fluorescence of air in high atmosphere conditions will help to improve the precision of the measure of the primary particle energy of cosmic rays\cite{AUGER}. Another purpose sought by PHIL is the training of the engineers, technicians and students that the accelerator community will need in the near future for other accelerator projects involving photoinjectors, like THOMX \cite{THOMX} for example.

\begin{figure}[h]
 \includegraphics[scale=0.5]{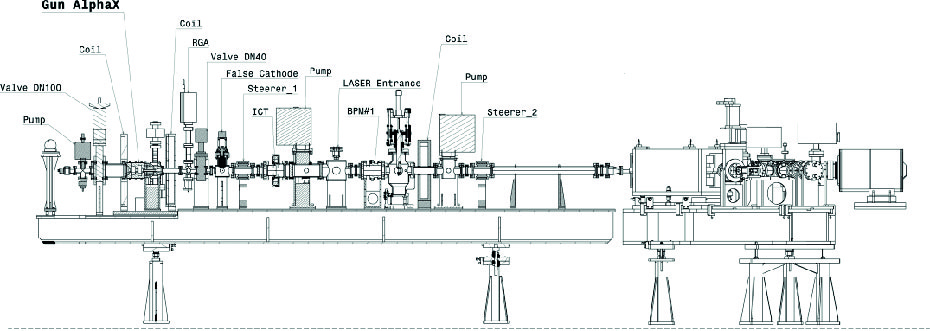}
                    \caption{Mechanical drawing of PHIL test line}\label{plan_meca}
\end{figure}

PHIL is a photoinjector beamline (see fig. \ref{plan_meca}), whose configuration evolves in time. Today PHIL is equipped with a copper photocathode RF gun, vacuum chambers with ionic pumps and magnetic elements: 
2 coils on the RF gun, 2 steerers to correct the orbit, a coil in the middle of the beamline and a dipole used as a spectrometer to measure the energy of the beam. The transverse positions of the electron beam are measured with Beam Position Monitors (BPM).  The charge is measured with Faraday cups at both ends of the beamline and
also with two Integrating Current Transformers (ICT from the Berghoz company). This setup is allowing us to measure the transport line charge transmission. The transverse sizes of the electron beam are measured with the light emitted by YAG:Ce screen stations coupled to CCD cameras \cite{refdiag}.

This article will describe the PHIL photoinjector beam line which is made up of 3 GHz RF gun,
a laser system, and a beam line diagnostics.

\section{PHIL RF guns and their sub-systems}

\begin{figure}[h]
  \centering
  \includegraphics[scale=0.5]{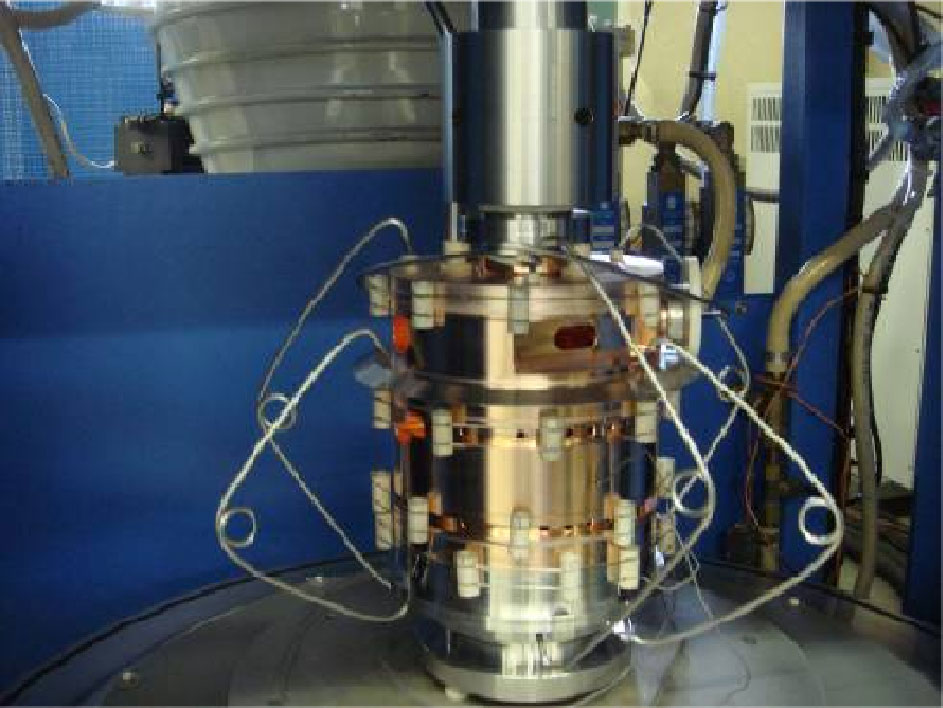}
                    \caption{picture of cells of the PHIN gun before their brazing under vacuum. The wires and the small ceramic tubes ensure that the cells are well centered. The load at the top and the springs guarantee the contact between surfaces which will be brazed. }\label{Canon_PHIN_brasage}
\end{figure}

At the beginning of the project PHIL, in 2004, the goal was to test the PHIN RF guns that LAL was manufacturing for CERN.
It was carried out as part of a work package of a joint research activity in the network CARE of the 6th Framework Program of the European Union. In addition to the 
construction of a photo-injector for CERN, LAL build its own beamline to do research and development on photoinjector technology.

These two RF guns have been totally built by the mechanical department of LAL. The design was performed jointly with the CERN 
experts and the workshop achieved the precise tolerances that are needed to fulfill both the RF and the brazing under vacuum constraints. 
A big effort was also produced to understand and optimise the different thermal cycles. This was needed to guarantee a successful
 brazing under vacuum. In picture \ref{Canon_PHIN_brasage}, you can see the PHIN gun, foreseen to be installed at PHIL, before one of 
the five brazing steps. 

The alphaX-RF gun was installed to start with and this allowed us to get a first electron beam at the end of 2009. 
The characterization of the beam produced is still ongoing. The drawback of the alphax RF gun is that, its cathode cannot be changed without a vacuum intervention. The next RF gun to be mounted on PHIL, will be
the PHIN RF gun which has the advantage of quick cathode changing without any vacuum breaking. Behind the RF gun will be placed a cathode holder coupled 
itself to four cathode chamber receiver. This last chamber can be coupled on the CERN preparation cathode laboratory, so that
different cathodes can be evaporated, transfered to Orsay, and tested on PHIL.   

\begin{figure}[h]
  \centering
  \includegraphics[scale=0.5]{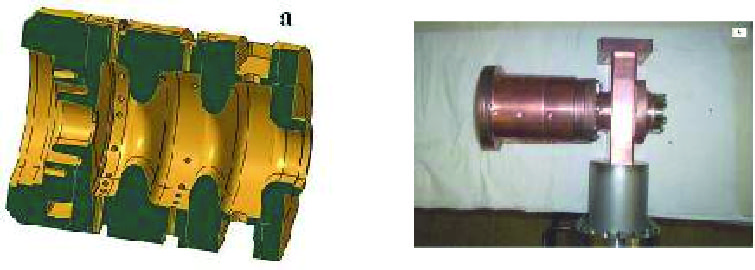}
                    \caption{a) PHIN gun, b) AlphaX gun}\label{photo-canon}
\end{figure}
\subsection{PHIN gun}
This photoinjector was designed to be the source of electrons for the drive beam linac of the CLIC Test 
Facility 3 (CTF3) at CERN \cite{ref1}. The specifications of CERN for this gun were rather demanding: produce 
around 2300 bunches of 2.33 nC each during the RF pulse of 2.5 $\mu$s with emittance below 20  $\pi$mmmrad 
and energy spread below 2 \%. It was decided that the design should rely on the long past experience of 
CERN in producing a high charge electron beam with photo-injector \cite{ref2}. The photo-injector has 2.6 cells
 at 2.998 GHz of resonant frequency, a schematic is shown in figure \ref{photo-canon}. The diameter of inner 
irises is rather large, 40 mm, to accommodate the high charge. The RF features are summarized in table 
\ref{table_canon}. Irises were machined with an elliptical shape which reduces the surface electrical field 
by 20 \% with respect to a cylindrical shape according to the RF simulations \cite{ref3, ref7}. 
In addition waveguides are symmetrically connected to the last cavity of the gun with respect to the mechanical
 axis in order to reduce emittance degradation induced by a non isotropic electrical field. 
The dimensions of the coupling apertures have been calculated in order to be at the critical coupling taking into account the strong beam loading induced by the electron beam, which average current in the drive beam linac is 3.5A 
Without beam, the coupling is roughly 3, divided into equal parts on both input waveguides. On this last point the operation will 
be drastically different on PHIL. Indeed for reasons of cost the laser of PHIL can deliver a single pulse at a repetition rate ranging from 5 to 100Hz.
Therefore there will be no beamloading which means that the PHIN gun, on PHIL, will stay overcoupled. In order to 
decrease the reflection factor we decided to close one input port with a short circuit which allows us to keep the 
symmetry of the electrical field in the gun while reducing the reflected power by a factor 4.

\subsection{AlphaX}

This gun has been installed on PHIL since 2009 and the first electron beam measurements were made with it. 
It is a copy of the photo-injector built by LAL for the ALPHAX accelerator in the University of 
Strathclyde in the UK \cite{STRAT}. The design of this gun was done by the Eindhoven University of Technology. 
It is also made of 2.5 cells at 2.998 GHz in the $\pi$ mode, a picture is shown in figure \ref{photo-canon} 
and also has elliptical irises. However the aperture of the irises is smaller than in the PHIN gun, 24 mm 
instead of 40 mm because the requirement of the extracted charge for AlphaX was 100 pC. But the main difference 
with respect to the PHIN gun is the coupling between the gun and waveguides which is done by a co-axial 
"doorknob" antenna in the cut-off tube after the cells of the gun. In this way the gun keeps a perfect cylindrical 
symmetry in order to avoid possible degradation of the emittance rising in non-symmetric coupling. RF characteristics
 are summarized in table \ref{table_canon}.

\begin{table}[h]
		\centering
		\begin{tabular}{||c||c||c||}
\hline
& PHIN & AlphaX \\ \hline
R$_s$ (M$\Omega$/m)	& 34	& 34 \\ \hline
Q	& 14530	& 11010\\ \hline
$\beta$ & 1.5	& 1 \\ \hline 	
		\end{tabular}
	\caption{RF characteristics of the RF guns used in PHIL; R$_s$ is the shunt impedance, Q the quality 
factor and $\beta$ the coupling factor.}
	\label{table_canon}
\end{table}

\subsection{RF power source}
The RF power needed for the gun is produced by a 25MW klystron (Thomson F2040E). 
LAL has built a classical Pulse Forming Network modulator for the 
biasing of this klystron. This modulator is made of an industrial high voltage supply 
(Technix SR20 20kV-0.4A) charging a Pulse Form Network (PFN) of 12 cells allowing a pulse length of 5 $\mu s$. 
The PFN is switched to a high voltage transformer (Stangeness) using a thyratron (E2V 1525AWX). 
This system is able to deliver pulses of 240kV/5$\mu$s/5Hz at the cathode of the klystron. A low power RF signal (350W/3$\mu$s, Nucletudes
 pre-amplifier) drives the klystron, the amplitude of the output RF power is modulated by variation of this low level signal. The klystron 
is protected from power reflection by a four port phase shift circulator. A standard WR284 waveguide network propagates the power to the gun.
The RF power is monitored at the output of the klystron and at the gun using two bi-directional couplers and crystal detectors. 
The wave guide network is under a 2 bars pressure of SF6, and RF windows isolate the klystron and the RF gun from the SF6. 

\begin{figure}[h]
  \centering
  \includegraphics[scale=0.5]{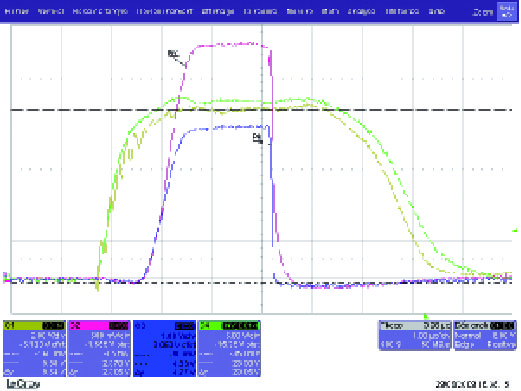}
        \caption{Measurement of Ik klystron intensity (yellow), Uk klystron voltage (green), Pik output power of the klystron (blue), Prk reflected power at the klystron (pink). The klystron was delivering 13MW}\label{rf}
\end{figure}

Power couplers at the exit of the klystron, and just before the RF gun allows a measurement of the different signals (see fig. \ref{rf}), as
the power going out of the klystron (Pik), the reflected power to the klystron (Prk), the power going inside the RF gun (Pic)
and the reflected power from the RF gun (Prc). Measurement of the pulse to pulse stability at 5 Hz have shown that it is better than $2~10^{-3}$

\subsection{Laser}
A Nd:YLF picosecond mode locked laser, model IC-262-40ps from `High Q Laser` company, 
is used to illuminate the photocathode. It consists of a SESAM passively mode locked 
picosecond oscillator, a regenerative amplifier and two frequencies converters and provides 
~80$\mu J$+/- 0.3$\mu J$ pulse energy ($\lambda=266nm$) at 9ps pulses with a pulse repetition frequency up to 100Hz.

\begin{figure}[h]
  \centering
  \includegraphics[scale=0.7]{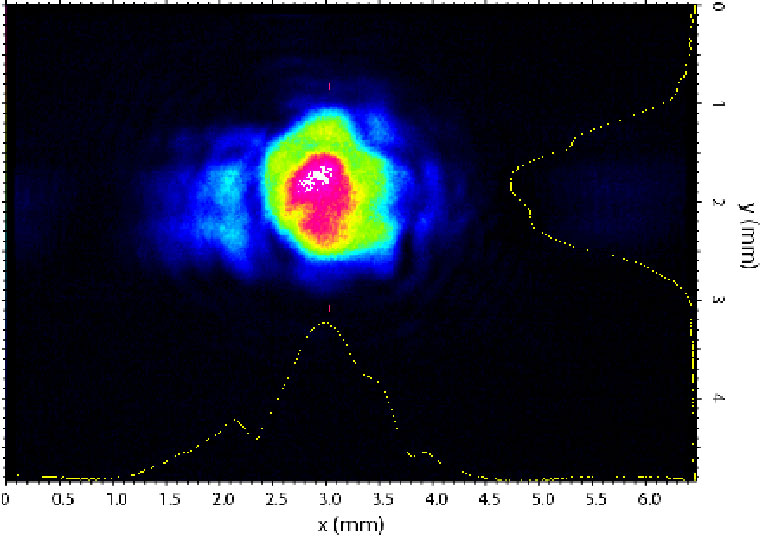}
                    \caption{Image taken with a UV CCD camera. The laser spot is observed on the virtual cathode on the 4th harmonic of the laser at 262 nm}\label{laser_spot}
\end{figure}

The laser oscillator can be locked to the 
external reference clock (frequency ~75MHz) coming from the pilot with the trigger better than 1ps (RMS).
The laser pulse to pulse energy stability is near 1$\%$ for roughly 8 hours. 
The output beam is shaped with the help of the iris aperture, and collimated by convex lens 
(2m-focal). After going through the optical transport line (~15m enclosed within a black plastic tube)
 the beam is focalized on the cathode by the 2m-focal lens at nearly normal incidence on the
 photocathode plane. The diameter of the laser spot on the photocathode is zoomed by means of 
the shifting of focusing lens and variation of the iris diameter. The spot size is controlled
 through an image of the cathode position and has a diameter about 1mm (see fig. \ref{laser_spot}). 
The beam point stability (standard deviation) on the cathode is 40$\mu m$. 

A streak camera (ARP) is used to check the pulse duration by sending the 266 nm attenuated light
directly to the camera. 

In the future, developements on the laser will include, obtaining a flat top transverse energy distribution,
a simple telescope system to vary the beam diameter continuously, and a stabilization system of the laser transverse 
position. The image of the laser will be captured with a dedicated UV camera. A feedback system will be installed based on the monitoring of the beam centroid position with a 4-quadrants diode to move accordingly the mirors with micrometric translators. 

\subsection{Timing}
\begin{figure}[h]
  \centering
  \includegraphics[scale=0.4]{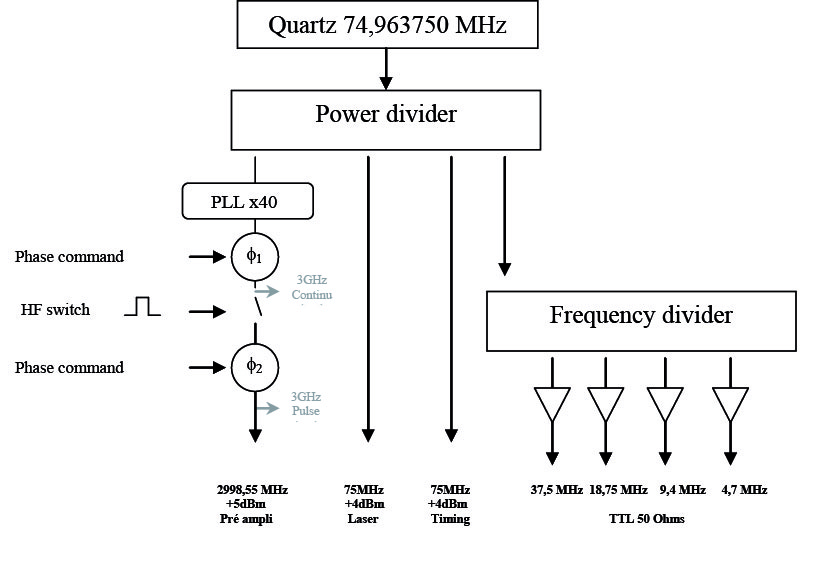}
                    \caption{Principle of the RF pilot}\label{rfpilot}
\end{figure}

Three main frequencies are used : 3 GHz for the RF wave, 5 Hz for the election bunch production and 75 MHz for the synchronisation signal. All the signals used on the accelerator are synchronised with a master-clock provided by the RF pilot (see fig. \ref{rfpilot}). 

This master clock has a frequency of 75MHz and is created
by an Oven Control Crystal Oscillator (OCCO), whose stability is $10^{-7}$. 
A PLL (Phase Locked Loop) is locked
on the master clock to create the low level RF signal (LLRF) at
the frequency of 2998.55 MHz, used for the accelerating structures.
 The phase of the LLRF is controlled by two adjustable phase-shifters.

The timing electronic system delivers all the slow signals
 synchronised on the master clock. Several PCI6602 National Instruments cards create
 the different TTL timing signals which trigger all the elements of the accelerator 
(RF pulse, laser pulse, diagnostics). For example, one timing signal triggers an RF 
switch to modulate the LLRF and so create the RF pulse that will drive the power source. 
Another one monitores the laser pulse at 75 MHz. All these signals have a repetition 
rate of 5Hz, their duration can be selected by the users from the command control computer.

\subsection{Vacuum}
\begin{figure}[h]
  \centering
  \includegraphics[scale=0.2]{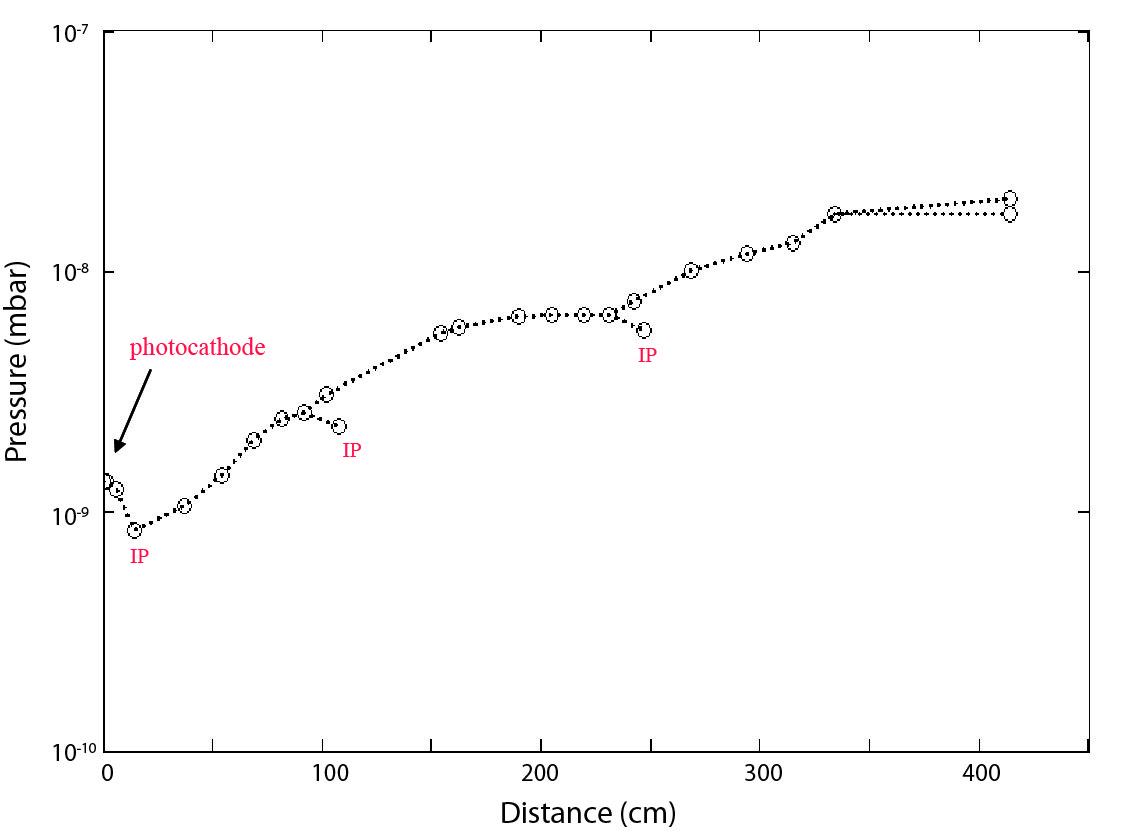}
                    \caption{Pressure distribution without in situ baking. Outgassing rate $\tau$=5~10$^{-11}$ mbar~l~s$^{-1}$~cm$^{-2}$ (eq. N$_2$)}\label{vide}
\end{figure}

One of the aims of PHIL is to use and test different photocathodes. Although, some alkaline photocathodes
 needs an ultra high vacuum (i.e. few 10$^{-10}$ mbar) to increase its lifetime. Then, the pumping 
distribution was calculated (see fig. \ref{vide}) to take into account that the downstream outgassing
 sources disturb the pressure minimally in the first half cell of the gun. For that purpose, ionic pumps (IP) were installed on the waveguides (2 IP starcell at 65L/s) and at the gun exit (1 IP starcell at 34L/s).
 A particular precaution was brought to the materials choices and to their thermal treatments. The PHIL components underwent UHV cleaning and vacuum baking of 450$^{\circ}$ C at 10$^{-6}$mbar for five days. At present, the pressure obtained at the level of the copper cathode of the alphaX gun is satisfactory in 2~10$^{-9}$mbar without in situ baking. The dynamic pressure remains suitable and increases by a factor 4 for a gun working at the temperature of 41$^{\circ}$ C. By cons, for the PHIN gun, which will receive alkaline photocathode, NEG coating was done on the gun output chamber to further improve the vacuum level. This will allow us to reach the 2~10$^{-10}$mbar level. 
Along the test line, ionic pumps were installed on the transport line (2 IP starcell at 125L/s) and upstream to every beam dump (1 IP starcell at 50L /s).

\subsection{Control systems}
\begin{figure}[h]
  \centering
  \includegraphics[scale=0.5]{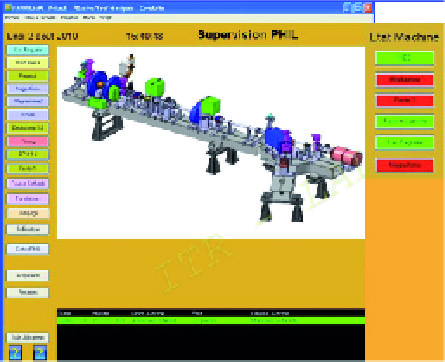}
                    \caption{The homepage of PHIL supervision (with Panorama software)}\label{Supervision}
\end{figure}

\begin{figure}[h]
  \centering
  \includegraphics[scale=0.55]{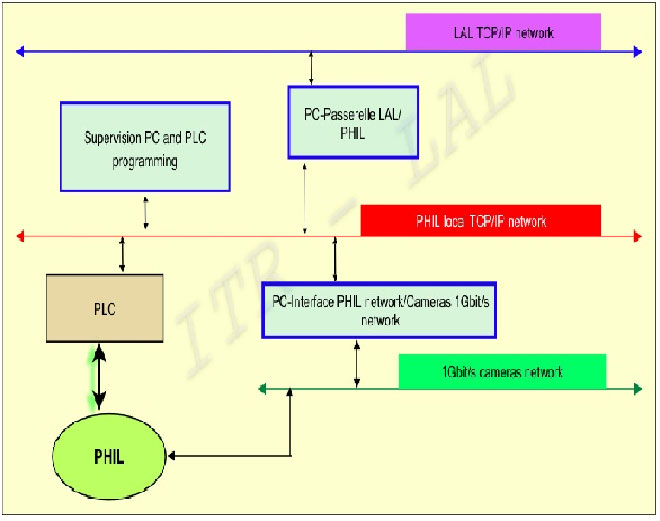}
                    \caption{Architecture of the PHIL control command }\label{Architecture}
\end{figure}
PHIL Command Control is centered on the monitoring soft
ware PANORAMA$^{TM}$ (see fig. \ref{Supervision}) of the company CODRA under
Windows XP and Ethernet private network, dedicated to the
Command Control of PHIL. The Architecture is illustrated in
figure \ref{Architecture}. The interface between the supervision and control
actuators and / or control is performed by PLCs (Wago) and
industrial PCs. The communication protocol is the Modbus
that through OPC servers can exchange the orders of Control and read or write variables to Control. Supervision also allows connection to an MSSQL database (Microsoft) to store recoverable acquisitions from the laboratory network. This allows
a web application to have a representation of the vacuum state
for example (only from the laboratory network) or the state of
the machine (accessible from the global network).

\subsection{RF commissioning of the guns}
Nowadays, the two guns have been conditionned : AlphaX up to 92MV/m and partially PHIN to 40MV/m. 

\begin{figure}[h]
  \centering
  \includegraphics[scale=0.4]{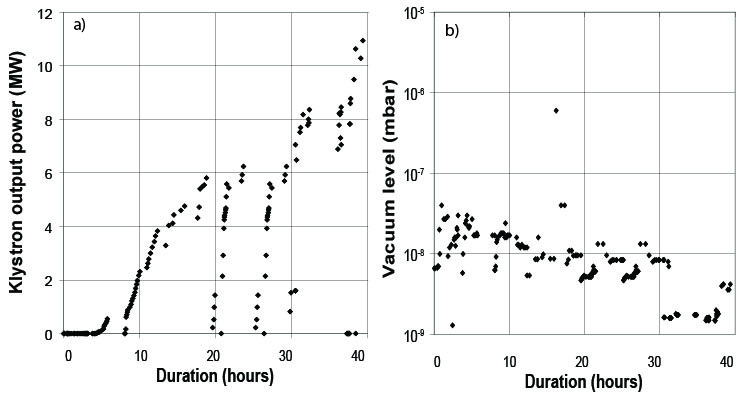}
                    \caption{a) Klystron output power versus time and b) gun vacuum level versus time during conditioning of the photoinjector AlphaX}\label{conditionnement}
\end{figure}

For five days (around 40h RF), the AlphaX RF commissioning was achieved in two phases : at first the conditionning was done just to be able to produce
a first low energy beam, with an input power of 4 MW inside the gun, in November 2009. 
Then the gradient inside the gun was increased on the beginning of 2010, and the gradient limit was tested. The incident power from the klystron was increased slowly, the vacuum and reflected power (see fig. \ref{conditionnement}a) of the
 gun were monitored. Some breakdowns occurred, but without damage, the vacuum level in the gun remained below 5.10$^{-8}$mbar during most of the process (see fig. \ref{conditionnement}b). Finally the klystron reached 11MW, corresponding to 10MW in the gun.
Taking into account the RF losses and the coupling we suceeded in injecting 10 MW in the alphaX RF gun, which corresponds to an accelerating gradient of 92 MV/m. To reach higher values, we have taken into account an adiabatic approach in the conditioning procedure (time consuming).
During the conditioning of the gun the vacuum composition was analysed with an RGA quadrupole placed as close as possible to the gun exit. It showed that gases emitted inside the gun are mostly CO$_2$ and H$_2$O.

\section{Magnetic elements}

In order to optimize beam quality and transport of the beam along the line,
magnetic elements were installed. PHIL beamline consists
of the three solenoids, two around the RF Gun, one at 2.266m from
the cathode plane, two steerers and one dipole.

The two solenoids around the gun are a bucking coil and a focusing coil. The former cancels the magnetic field in the photo-cathode plane and the other is used to focus the particles
 at the exit of the gun. The positions of the coils have been calculated and optimized to limit the emittance growth due to space charge forces. The magnetic field calculated to keep this low emittance value is 0.25T at 2nC and $E_{acc}$=85MV/m. The design of the coils was optimized for the PHIN gun geometry. It involved constraints on the dimensions of the coil. First, along
the longitudinal axis, the minimum distance between the bunching coil and the focusing one could only be 11cm. Secondly, due to the presence of a vacuum chamber around the RF gun,
 the inner diameter of the coils is rather large. It is then necessary to optimize the number of windings, the diameters and the current in order to stay within the limits imposed by the material
 while fulfilling the specification to reach a maximum magnetic
field of 0.28T. The solenoid were manufactured by a private company SEF. The conductor used for windings is a square copper cable, 6x6 $mm^2$ with a hollow in the center of 4mm diameter for water cooling. The focusing coil has 208 windings and the other one 100. The Brucker power supply can produce a maximal current of 400A allowing
one to reach the magnetic field of 0.28T. 

\begin{figure}[h]
  \centering
  \includegraphics[scale=0.4]{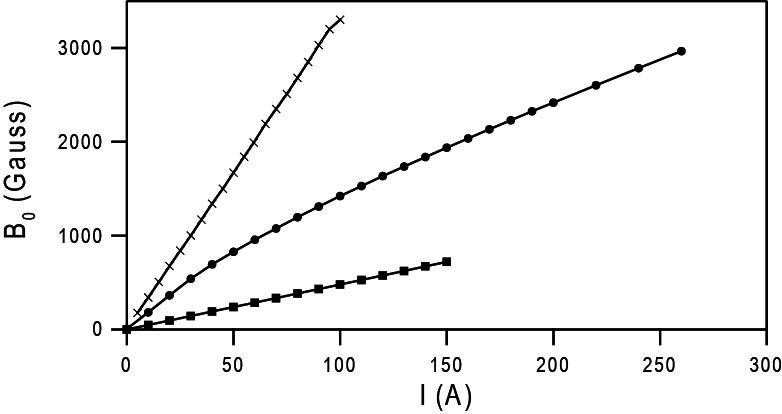}
                    \caption{Measurement of the maximum on axis magnetic field versus current for the three solenoids of PHIL line: circle) emittance compensation coil at the exit of the gun, square) bucking coil and cross) focusing coil. The emittance compensation coil exhibits a non linearity resulting in a variation of its magnetic length depending on the current.}\label{bobines}
\end{figure}

However let's  note that up to now, it is the ALPHAX RF gun that was installed on PHIL. As this rf gun has a different geometry, the magnetic configuration is different as, for example, the distance between coils. The cancelation of the solenoidal field at the cathode plane \cite{ref5}
is possible with this gun but with a different set up than PHIN gun. The last solenoid located in the middle of the beam line is also used to focus particles, like a quadrupole triplet would have done. But looking at the energy range and the cost, a solenoid fulfills all the requirements needed. This solenoid is able to reach a maximum magnetic field of 0.5T. It has an inner diameter of 70mm and 208 windings. The measured calibration of the three solenoids is illustrated in figure \ref{bobines}. The emittance compensation coil exhibits a non linearity \cite{ref4}, which implies variation of its magnetic length with the magnetic field amplitude. 

The two steerers are used to correct the orbit. One is located roughly 70cm after the output of the gun and one just before the dipole. Technically, they consists of 40 turns in each coil allowing a maximum magnetic field of 20G for an intensity of 5A. The power supply of this magnet is bipolar going from -5A to 5A to allow bending in opposite directions. With these steerers, the trajectory can be changed up to 10mrad.

Concerning the dipole, it was made originally for TTF injector \cite{Aimant1} and was used with a maximum magnetic field of 0.1T for 20A. This dipole is now used as a spectrometer to measure the beam energy. This is a C-dipole (with face angle of 18.24°) with a curvature radius of 0.7m and a bending angle of 60$^{\circ}$. Magnetic measurements were done at CERN to map the magnetic field. This dipole reaches a magnetic field
of 421.5G for 8.6A corresponding to the maximum energy of 10MeV.

\section{Diagnostics}

\subsection{Electron beam energy and its dispersion}
\begin{figure}[h]
  \centering
  \includegraphics[scale=0.6]{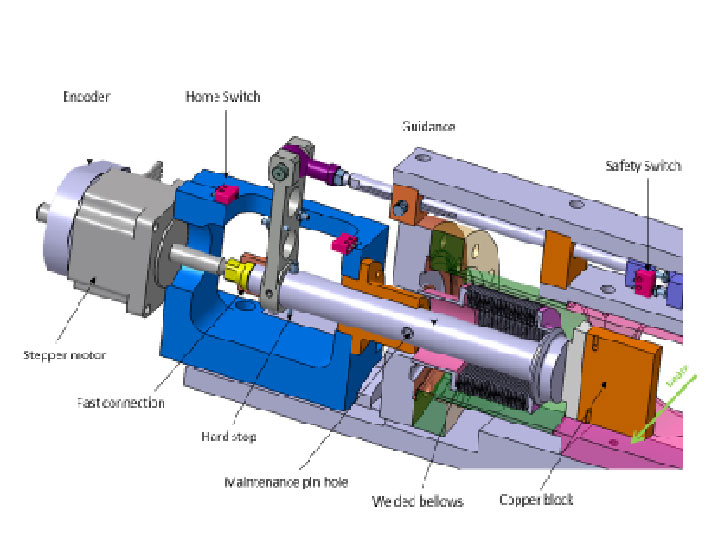}
                    \caption{3D view of the energy spread collimator. The copper block is actuated by a stepper and guided with the axle. The position is read by the encoder. }\label{fente}
\end{figure}

On the PHIL Beamline, the dipole is used as a spectrometer
to measure the energy of the beam. Placed after the dipole a
collimator is used to analyse the energy distribution. 
The collimator is made of 2 copper blocks actuated separately at 90$^{\circ}$
from the beam axis. This produces a slit with an adjustable size
(from 1mm to 50mm) and position (+/- 25mm from the beam
axis). A stepper motor and a precise angular encoder were used
to achieve the required accuracy (+/-0.1mm in slit dimensions
and +/-0.1mm in position) (see picture \ref{fente}). Measurements and
load tests were performed before the fabrication of the device.
This collimator was fabricated in the LAL workshops.

\begin{figure}[h]
  \centering
  \includegraphics[scale=0.4]{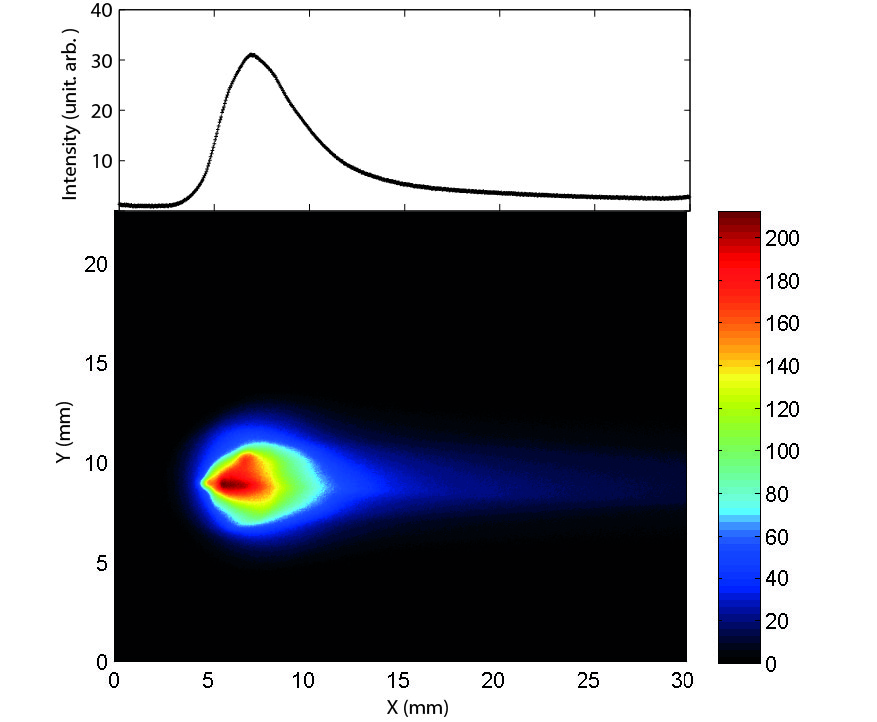}
                    \caption{Visualisation of the transverse projection of the electron beam on the YAG:Ce screen located after the dipole in the dispersive region. The projected horizontal distribution is represented on the top of the image.}\label{dispersion}
\end{figure}

The electron beam energy dispersion is directly related to the size of
the beam by the dispersion function, in the focal plane of the dipole. 
Measuring the beam size leads to measurement of the energy dispersion as a YAG:Ce screen is placed close to the dipole focal plane. An example is shown in figure \ref{dispersion}.

\subsection{Charge}
\begin{figure}[h]
  \centering
  \includegraphics[scale=0.4]{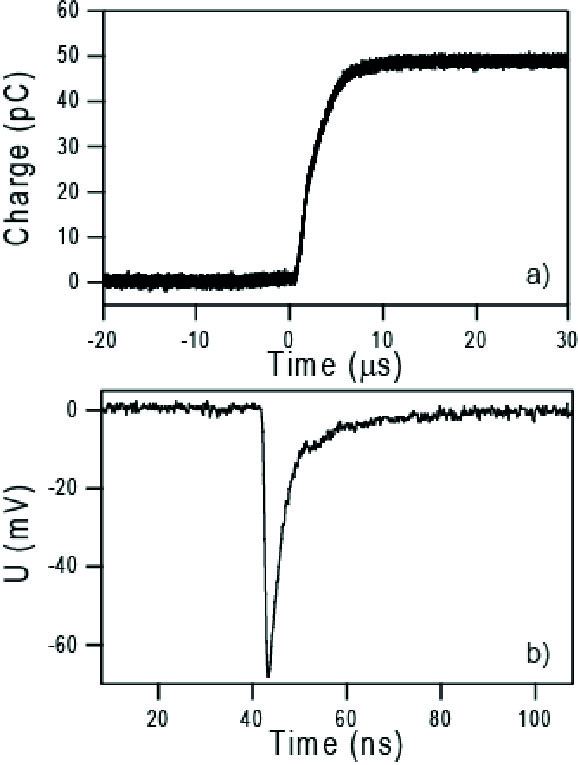}
  \caption{Signal of the a) ICT , b) Faraday cup.}\label{charge}
\end{figure}

The charge is measured (see fig. \ref{charge}) with Faraday cups at both ends of the
beamline and also with two Integrating Current Transformer installed in front of the Faraday cup at the end of the straight beam line and one just after the gun. 

The extracted charge is a key parameter to control for 
some applications like detector calibration for example.
It is measured simultaneously with a Faraday cup and an ICT (less than 10 pC to 2nC) at
the end of the beam line, these two devices have been calibrated
and are in good agreement. First experiments give an extracting
charge of 500 pC (+/- 10pC) with a laser pulse energy of 50 $\mu J$
(with a fluctuation of 0.5 $\mu J$). 

\begin{figure}[h]
  \centering
  \includegraphics[scale=0.6]{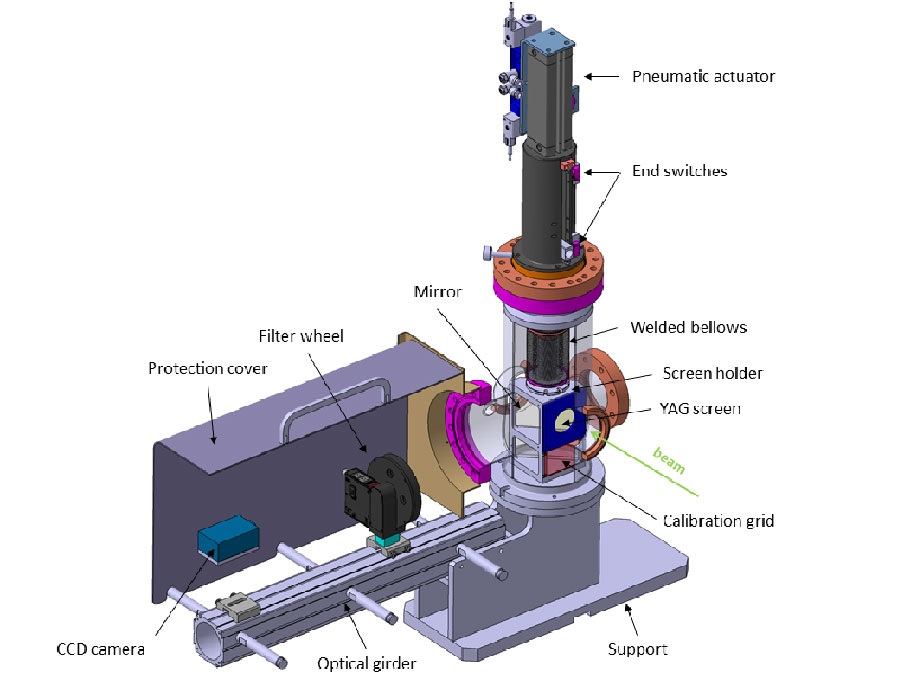}
  \caption{3D view of the new screen station. The YAG:Ce screen at 90° from the beam is inserted on beam axis with the pneumatic actuator. The profile, reflected with a mirror at 45°, is analysed by a CCD camera at the end of the optical girder. }\label{Profileur_phil_2}
\end{figure}

\subsection{Transverse dimensions}
The transverse sizes of the electron beam are measured with
 a YAG:Ce screen (YAG1) coupled to a CCD camera placed just before
 the coil in middle of the beam line (see fig. \ref{plan_meca}). The resolution of the optical system (made of one achromatic doublet) has been estimated to be about 80$\mu m$. This resolution level is mainly induced by the orientation (45$^{\circ}$) and the thickness (300$\mu m$) of the YAG:Ce screen \cite{X1}. A remote control filter wheel (equipped with discrete variable Optical Density) is placed just in front of the CCD in order to avoid image saturation. The cameras used are Blue Cougar Matrix Vision CCD camera type S123 (resp. S120) pixel size of 1360x1024 (resp. 650x490).

A MATLAB code has been written in order to analyse the beam profile. A gaussian fit on the projected horizontal and vertical distributions enables to extract the diameter of the beam\cite{X2}. An online extraction data HIM (based on Labview) is under development and will be operationnal in fews months \cite{HIM}. Also a comparison of the fit gaussian method and the moment distribution method is in progress.
 
Three other YAG:Ce screens were installed in 2011 : one just before the dipole, a second at the end of the straight beam line and a third on the deviated beam line after the dipole. The optical system used for these stations are slightly different from the YAG1 station. The screen orientation is at 90$^{\circ}$ for YAG2, 3 and 4, instead of 45$^{\circ}$ for YAG1. A better resolution should be obtained with this set up. The light emitted by the YAG:Ce screen goes after a specular reflexion through an achromatic lenses doublet
on the CCD pixel chip. The new YAG:Ce screen stations (see fig. \ref{Profileur_phil_2}) are composed of a screen holder with a YAG:Ce screen perpendicular to the beam direction, a mirror placed at 45$^{\circ}$ 
and a calibration grid below, moved vertically with a two-position pneumatic jack, for checking
the calibartion pixels. We used only a simple guidance axis in order to have a basic, cheap and reliable system. The choice of avoiding the mechanical adjustments led us to have precise interfaces between the PHIL girder and the screen guidance but simplify the mounting and the operation 
of these screen stations (e.g. mirror or screen exchange).

\begin{figure}[ht]
  \centering
  \includegraphics[scale=0.5]{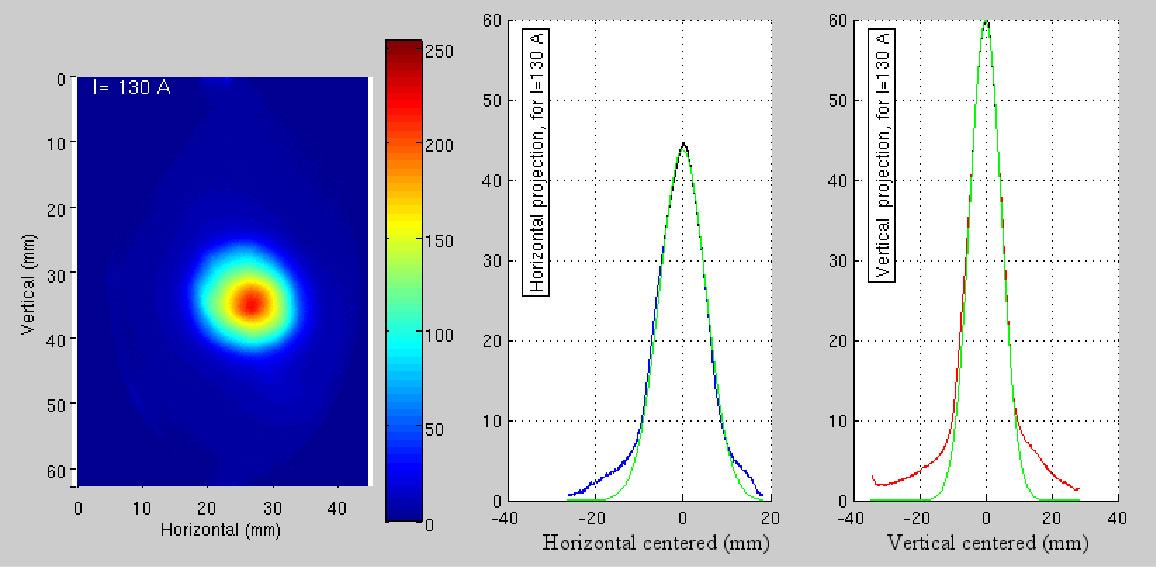}
  \caption{Left panel : transverse beam profile obtained on the YAG:Ce screen at 1925 mm from the
photocathode, with a focusing solenoid (located at the exit of the Alphax gun) alimented by 130 A. Right panel : horizontal and vertical beam projection with gaussian fit (in green).}\label{profil_trans}
\end{figure}

\subsection{Bunch length}
The bunch duration of the electron beam will be measured
using a streak camera analysing the Cerenkov light emitted
by the electrons crossing a thin saphir plate. The Cerenkov
radiator is placed inside the beam pipe under vacuum. The  
Cerenkov light is transported over 15 m in air, inside a tubed
path, up to the ARP streak camera, with metallic mirors and achromatic
lenses. We expect a rate of five photons per electron between 400 and 600 nm for a 5 MeV electron beam energy and 200$\mu m$ saphir thickness. This is sufficient to have a good signal detected by the streak camera. The resolution of our streak camera is 3ps. The complete set up of the bunch length is under installation and will be available by 2013.

\subsection{Transverse emittance}
PHIL will be equipped with a two-dimensional transverse beam emittance measurement system based on the multi-slits method.
This technique - briefly described here - is realized by introducing a slits mask in the beam trajectory, and by observing the output
beamlets at a downstream location. The position and the thickness of the beamlets allows to reconstruct the 2D trace space
$(x,x')$ or $(y,y')$ and the geometrical 2D rms transverse emittance \cite{A}. The accuracy of the measurement depends on the space charge effect, and all the parameters of the system should be carefully chosen for a given range of beam setup \cite{B}.
For PHIL (E$<$10MeV), a detailed analysis \cite{C} has been carried out using a home-made Matlab code \cite{D}.
This study fixed the following values for the system \cite{E}:
\begin{itemize}
	\item mask material : tungsten (W) 
\item thickness of the mask : 3.5 mm
\item number of slits : 27
\item thickness of the slits : 0.1 mm
\item distance between 2 slits : 1.5 mm
\item distance between slits and screen : 230 mm
\end{itemize}
First measurement using this system are planned for 2013.

\section{Conclusion}
PHIL has commisionned several elements and can deliver electron beams routinely. Other elements like the bunch length and emittance station measurement are expected to fully caracterised the RF guns. A full characterization of the beam, under different configurations will be done
this year. The understanding for sources of instabilities avoiding reproductibility is under characterisation in order to be able to develop feedback systems. A first user experiment is expected during the year 2012.

\section{Acknowledgments}
We would like here to acknowledge the support of all people helping PHIL in a many ways starting with the scientific committee of PHIL. The collaboration with CERN/CTF3 PHIN photoinjector team also brings a lot to the PHIL developpement. Ideas given by the PITZ Zeuthen team are also very helpful.
We also thank the local radioprotection team without wich PHIL would not be able to run. The support of the IN2P3 (Institut National de Physique Nucleaire et des Particules) and P2IO (Physique des deux infinis) is also acknowledged here. The authors also acknoledge G. Bienvenu, B. Mouton, B. Leblond, who initiated the design of PHIL during CARE.








\end{document}